\documentclass{aa}
\usepackage{amsmath}
\usepackage{txfonts}
\usepackage{graphicx,float,subfigure}
\usepackage{natbib}
\bibpunct{(}{)}{;}{a}{}{,}
\newcommand{\dd}{\ensuremath{\mathrm{d}}}
\newcommand{\DD}{\ensuremath{\mathrm{D}}}

\begin{document}

\title{A localised subgrid scale model for fluid dynamical simulations in
astrophysics II: Application to type Ia supernovae\\ }

\author{W. Schmidt}
\author{W. Schmidt\inst{1,2} \and J.~C. Niemeyer\inst{1} \and W. Hillebrandt\inst{2} \and F.~K. R\"{o}pke\inst{2}}

\institute{Lehrstuhl f\"{u}r Astronomie, Institut f\"{u}r Theoretische Physik
und Astrophysik, Universit\"{a}t W\"{u}rzburg, Am Hubland, D-97074
W\"{u}rzburg, Germany \and Max-Planck-Institut f\"{u}r Astrophysik,
Karl-Schwarzschild-Str.\ 1,\\ D-85741 Garching, Germany}

\date{Received / Accepted}

\titlerunning{A localised subgrid scale model for fluid dynamical simulations II}
\authorrunning{W. Schmidt et al.}

\abstract{ The dynamics of the explosive burning process is highly sensitive to the
	flame speed model in numerical simulations of type Ia
	supernovae.  Based upon the hypothesis that the
	effective flame speed is determined by the unresolved
	turbulent velocity fluctuations, we employ a new subgrid scale
	model which includes a localised treatment of the energy
	transfer through the turbulence cascade in combination with semi-statistical
	closures for the dissipation and non-local transport of
	turbulence energy. In addition, subgrid scale buoyancy effects
	are included. In the limit of negligible energy transfer and
	transport, the dynamical model reduces to the Sharp-Wheeler
	relation. According to our findings, the Sharp-Wheeler relation
	is insuffcient to account for the complicated turbulent
	dynamics of flames in thermonuclear supernovae. The
	application of a co-moving grid technique enables us to
	achieve very high spatial resolution in the burning
	region. Turbulence is produced mostly at the flame surface and
	in the interior ash regions. Consequently, there is a
	pronounced anisotropy in the vicinity of the flame fronts. The
	localised subgrid scale model predicts significantly enhanced
	energy generation and less unburnt carbon and oxygen at low
	velocities compared to earlier simulations.
  
 }

\maketitle

\section{Introduction}

For supernovae of type Ia, \citet{HoyFow60} proposed a thermonuclear
runaway initiated in C+O white dwarfs close to the Chandrasekhar limit
as the cause of the explosion. Since the original proposal, there has
been vivid controversy of how such an explosion might come about and
what the exact physical mechanism could be. Today the computational
facilities to process three-dimensional large-eddy simulations (LES) of the
explosion event are available. Remarkably, these powerful means have
not aided in arriving at a consensus yet. The disagreement stems from
some crucial questions. Firstly, what is the appropriate flame
speed model?  Secondly, does the explosion completely proceed as
a deflagration, or does a transition to a delayed detonation set in at
some point?  The deflagration to detonation transition (DDT) proposed
by \citet{Khok91a} and \citet{WoosWeav94} appears to resolve the
drawbacks of the pure deflagration model.  In particular, the energy
output obtained from simulations with artificial DDT is closer to the
observed one, and less carbon and oxygen is left behind
\citep{GamKhok04,GamKhok05,GolNie05}.  For the theoretical
understanding of thermonuclear supernovae, however, the lack of a
convincing explanation for the initiation of the transition is
unsatisfactory \citep{KhokOr97,NieWoos97,Nie99,ZingWoos05}. In the
aforementioned numerical models, a DDT is artifically triggered at
more or less arbitrary instants of time.

As for the flame speed model, the controversy is whether subgrid scale
(SGS) turbulence is mostly driven by Rayleigh-Taylor instabilities or
dominated by the energy transfer through the turbulence cascade.
The former point of view holds that the magnitude of SGS
velocity fluctuations $v^{\prime}$ is basically given by the Sharp-Wheeler relation
\citep{DavTay50,Sharp84}
\begin{equation}
	\label{eq:vel_sharp}
	v_{\mathrm{RT}}(l)=0.5\sqrt{l g_{\mathrm{eff}}}
\end{equation}
where $v_{\mathrm{RT}}(l)$ is the asymptotic rise velocity of
a perturbation of size $l$ due to buoyancy. The effective gravity
$g_{\mathrm{eff}}$ is determined by the density contrast
at the interface between burned and unburned material.
Setting the turbulent flame speed equal to $v_{\mathrm{RT}}(\Delta)$,
where $\Delta$ is the resolution of the numerical grid,
has been used in some simulations of type Ia supernovae
\citep{GamKhok03,CaldPle03,GamKhok04}.
However, simple scaling arguments disfavour this proposition \citep{NieHille95a,NieKer97}.
Assuming that non-linear interactions between turbulent eddies of different
size $l$ set up a Kolmogorov spectrum, the root mean square turbulent velocity
fluctuations obey the scaling law $v^{\prime}(l)\propto l^{1/3}$.
Since the Sharp-Wheeler relation implies $v_{\mathrm{RT}}(l)\propto l^{1/2}$, we
have $v_{\mathrm{RT}}(l)/v^{\prime}(l)~\propto l^{1/6}\rightarrow 0$ 
towards decreasing length scales.
Consequently, \citet{NieHille95a} adopted a SGS model based
on the dynamical equation for the turbulence energy $k_{\mathrm{sgs}}$, i.e. the
kinetic energy of unresolved velocity fluctuations \citep{Schumann75}. 
The major weakness of their approach arises from the fairly tentative closures 
which were formulated \emph{ad hoc} for LES of stellar convection \citep{Clement93}. 

Various refutations of the scaling argument have been put forward.
To begin with, the spectrum of turbulence energy might be different
from the Kolmogorov spectrum. However, recent direct numerical
simulations support the hypothesis of a Kolmogorov spectrum
in buoyancy-driven turbulent combustion \citep{ZingWoos05}. A more serious
concern is that there might be not enough time to reach the
state of developed turbulence with a Kolmogorov spectrum
in the transient scenario of a supernova explosion. This
question is difficult to settle \emph{a priori}.  For this
reason, we took an unbiased point of view and accommodated
buoyancy effects in the form of an Archimedian force term
in the SGS turbulence energy model.

In contrast to the previously used SGS turbulence energy model, 
the new localised model which is thoroughly discussed in paper I
neither presumes isotropy nor a certain turbulence energy spectrum
function. This is possible by virtue of a dynamical
procedure for the determination of the SGS eddy-viscosity $\nu_{\mathrm{sgs}}=
C_{\nu}\Delta k_{\mathrm{sgs}}^{1/2}$ which was adapted
from \cite{KimMen99}. Furthermore, we apply the co-expanding
grid introduced by \cite{Roep05}. The growth of the cutoff length $\Delta$
due to the grid expansion poses a challenge for the SGS model because 
of the partitioning between resolved energy and SGS energy changes in time. 
We will show that this rescaling effect can be taken into account
by utilising the dynamical procedure for the calculation of eddy-viscosity parameter
$C_{\nu}$. The rescaling algorithm as well as the computation of the
Archimedian force is explained in Sect.~\ref{sc:flame_speed_model}, 
followed by the discussion of results from three-dimension numerical simulations
in Sect.~\ref{sc:num_simul}. It is demonstrated that the newly proposed
SGS model substantially alters the predictions of the deflagration model.
In particular, we will analyse the significance of SGS buoyancy affects.

\section{The flame speed model}
\label{sc:flame_speed_model}

For the relation between the turbulent flame speed $s_{\mathrm{t}}$
and the SGS turbulence velocity $q_{\mathrm{sgs}}$, we adhere to the results 
found by \citet{Poch94} from a theoretical analysis and set
\begin{equation}
  \label{eq:sgs_flame_speed_pocheau}
  s_{\mathrm{t}} = s_{\mathrm{lam}}
  \sqrt{1 + C_{\mathrm{t}}
        \left(\frac{q_{\mathrm{sgs}}}{s_{\mathrm{lam}}}\right)^{2}},
\end{equation}
where $C_{\mathrm{t}}=4/3$. In the asymptotic regime of turbulent
burning, $s_{\mathrm{t}}\simeq 2q_{\mathrm{sgs}}/\sqrt{3}$
which is consistent with \citet{Peters99}.

The evolution of $q_{\mathrm{sgs}}$ is given by a non-linear partial
differential equation which is obtained by dividing the equation
for the specific SGS turbulence energy 
$k_{\mathrm{sgs}}=\frac{1}{2}q_{\mathrm{sgs}}^{2}$ (see Sect.~3 of
part I) by $q_{\mathrm{sgs}}$. For turbulence driven
by the Rayleigh-Taylor instability, an additional
source term stems from buoyancy effects
on subgrid scales. This Archimedian force,
which is proportional to the effective gravity due
to the density contrast between nuclear fuel and ash,
directly induces turbulent velocity fluctuations. The
form of the Archimedian force will be proposed in Sect.~\ref{sc:archm_prod}.
Moreover, the Eulerian time derivative must account for the
rescaling of the turbulence energy due to the temporal shift
of the cutoff length. We will denote the partial derivative
with respect to the rescaled quantities by $\partial^{\star}$.
The Lagrangian time derivative thus becomes
\begin{equation}
  \frac{\DD^{\star}}{\DD t} = \frac{\partial^{\star}}{\partial t} + \vec{v}\cdot\vec{\nabla}.
\end{equation}

The complete dynamical equation for the SGS turbulent velocity is
\begin{equation}
\label{eq:q_sgs}
\begin{split}
  \frac{\DD^{\star}}{\DD t}q_{\mathrm{sgs}} 
  -& \frac{1}{\rho}\vec{\nabla}\cdot\left(\rho\ell_{\kappa}q_{\mathrm{sgs}}
     \vec{\nabla}q_{\mathrm{sgs}}\right) -
     \ell_{\kappa}|\vec{\nabla}q_{\mathrm{sgs}}|^{2} \\
  &= \frac{1}{\sqrt{2}}C_{\mathrm{A}}g_{\mathrm{eff}} +
     \ell_{\nu}|S^{\ast}|^{2} - \frac{7}{30}q_{\mathrm{sgs}}d -
     \frac{q_{\mathrm{sgs}}^{2}}{\ell_{\epsilon}}.
\end{split}
\end{equation}
The rate-of-strain scalar $|S^{\ast}|$ is defined by
$|S^{\ast}|^{2}=2S_{ij}^{\ast}S_{ij}^{\ast}$, where
$S_{ij}^{\ast}$ is the trace-free part of the
symmetrised Jacobian matrix of the velocity field,
$S_{ij}=\frac{1}{2}(\partial_{j}v_{i}+\partial_{i}v_{j})$,
and $d=S_{ii}=\partial_{i}v_{i}$ is the divergence.
The characteristic length scales $\ell_{\kappa}$,
$\ell_{\nu}$ and $\ell_{\epsilon}$ 
are related to SGS turbulent transport, the rate
of energy transfer from resolved toward subgrid scales
and the rate of viscous dissipation. Each characteristic
length can be expressed in terms of the effective
cutoff length $\Delta_{\mathrm{eff}}$ and a 
similarity parameter:
\begin{equation}
	\ell_{\nu}=\frac{C_{\nu}\Delta_{\mathrm{eff}}}{\sqrt{2}},\qquad
	\ell_{\epsilon}=\frac{2\sqrt{2}\Delta_{\mathrm{eff}}}{C_{\epsilon}},\qquad
	\ell_{\kappa}=\frac{C_{\kappa}\Delta_{\mathrm{eff}}}{\sqrt{2}}.
\end{equation}
For the determination of the closure parameters $C_{\nu}$, $C_{\epsilon}$
and $C_{\kappa}$ see Sect.~4 of part I. The advection of 
$q_{\mathrm{sgs}}$ by the resolved flow is computed
with the piece-wise parabolic method \citep{ColWood84}. Due to the dissipative
effects of this numerical scheme on the smallest resolved length
scales, we set $\Delta_{\mathrm{eff}}\approx 1.6\Delta$ \citep{SchmHille05a}.
The diffusion terms on the left hand side of equation~\ref{eq:q_sgs} is computed by means
of fourth order centred differences, and for the source term on the right-hand side a semi-implicit
Adams-Moulton method is used. In the remainder of this section, we describe the
calculation of the SGS Archimedian force and the rescaling procedure.

\subsection{Archimedian production}
\label{sc:archm_prod}

There has been an ongoing debate whether the
production of SGS turbulence is dominated by the buoyancy of
SGS perturbations in the interface between ash and fuel or
by eddies produced through the turbulence cascade. In the first case, the source
of energy is the gravitational potential energy, whereas non-linear
transfer supplies kinetic energy from larger scales in the second case.
In general, it is quite difficult to separate the energy injection caused by gravity in wave number space
because gravitational effects are genuinely non-local. For Rayleigh-Taylor driven
turbulence, however, we know the simple Sharp-Wheeler scaling relation~(\ref{eq:vel_sharp})
which provides an algebraic closure for the Archimedian force density $\Gamma_{\mathrm{sgs}}$
introduced in Sect.~3 of part I. Therefore, we propose a novel approach which combines both the production of
SGS turbulence through the cascade and the Rayleigh-Taylor mechanism in the dynamical
equation for $q_{\mathrm{sgs}}$.

Because $\Gamma_{\mathrm{sgs}}$ has the dimension of an acceleration times
velocity, we interpret this term as the product of a specific \emph{Archimedian force} and the
SGS turbulence velocity $q_{\mathrm{sgs}}$. This means the following: In any finite-volume
cell those portions of the fluid with density less than the smoothed density $\rho$ 
will experience buoyancy relative to the other portions of higher density in the
mean gravitational field. In subsonic turbulent flows, the small random density fluctuations
caused by compression and rarefaction are expected to produce very little net buoyancy. However,
a special situation is encountered in the cells intersected by the flame fronts.
Since the flames are far from being completely resolved in numerical simulations
of SNe Ia, the substructure of the front in combination with the density gradient
across the front will produce SGS buoyancy. Equivalently, one can think of of
perturbations in the flame front on length scales $\lambda_{\mathrm{fp}}\lesssim l\lesssim\Delta$
as being Rayleigh-Taylor unstable and producing SGS turbulence. The \emph{fire polishing length}
$\lambda_{\mathrm{fp}}$ then marks the lower threshold for perturbations to grow
\citep[see][]{ZingWoos05}. Once turbulence has developed, $\lambda_{\mathrm{fp}}$ 
can be identified with the Gibson length $l_{\mathrm{G}}$, i.e. the smallest
length scale on which the flame propagation is affected by turbulent eddies. Perturbations
of size $l\gtrsim\Delta$, on the other hand, set fluid into motion on numerically
resolved length scales. But the transfer
of energy through the turbulence cascade eventually produces SGS turbulence as well.
This production channel corresponds to the term $\Sigma_{\mathrm{sgs}}$ in the equation
for the SGS turbulence energy (see sect~3 of part I).

The Archimedian force generated by the density gradient across flame fronts
is given by the effective gravity
\begin{equation}
	g_{\mathrm{eff}}=\mathrm{At}\,g,
\end{equation}
where the Atwood number
\begin{equation}
	\mathrm{At} = \frac{\rho_{\mathrm{f}}-\rho_{\mathrm{b}}}
	                   {\rho_{\mathrm{f}}+\rho_{\mathrm{b}}}
\end{equation}
is a measure for the density contrast between burned material and nuclear fuel.
In the vicinity of the flame front, the lowest order estimate for the SGS
buoyancy term is $\Gamma_{\mathrm{sgs}}\sim\rho q_{\mathrm{sgs}}g_{\mathrm{eff}}$,
provided that $\Delta\ga\lambda_{\mathrm{fp}}$. If $\Delta$ becomes smaller than 
$\lambda{\mathrm{fp}}$, SGS perturbations in the flame front are not subject to the RT-instability
and $\Gamma_{\mathrm{sgs}}$ vanishes. In conclusion, we propose the following closure:
\begin{equation}
	\label{eq:buoy}
	\Gamma_{\mathrm{sgs}} =
	\frac{1}{\sqrt{2}}C_{\mathrm{A}}\rho g_{\mathrm{eff}}q_{\mathrm{sgs}}.
\end{equation}
Here the effective gravity is more precisely defined by
\begin{equation}
	g_{\mathrm{eff}} =
	\chi_{\pm\delta}(G=0)\theta(\Delta-\lambda_{\mathrm{fp}})
	\mathrm{At}(\rho)g,
\end{equation}
where $\chi_{\pm n\Delta}(G=0)$ is the characteristic function of
all cells for which the distance from the flame front (represented by
$G(\vec{x},t)=0$) is less than $\delta$, and $\theta$ is the Heaviside
step function, i.e. $\theta(\Delta-\lambda_{\mathrm{fp}})=1$ for
$\Delta>\lambda_{\mathrm{fp}}$ and zero otherwise. The Atwood number is
expressed as a function of the mean density which is obtained from
a fit to the numerical data from \citet{TimWoos92}:
\begin{equation}
    \mathrm{At}(\rho) = \frac{1}{2}\left[0.0522 + \frac{0.145}{\sqrt{\rho_{9}}}
      - \frac{0.0100}{\rho_{9}}\right]
\end{equation}

The closure~(\ref{eq:buoy}) does not include all of the intricate effects
that contribute to SGS buoyancy. It merely captures what is presumably the leading
order effect. In fact, one would have to model the interaction
between turbulent potential and kinetic energy fluctuations on unresolved scales.
Unfortunately, there exists no theoretical framework for this task yet. Moreover, the concept
of a SGS Archimedian force entails a violation of energy conservation because
the contribution of $\Gamma_{\mathrm{sgs}}$ to the production of turbulence energy
is not balanced in the resolved energy budget. Consequently, the total energy
of the system effectively increases.
However, $\Gamma_{\mathrm{sgs}}$ is non-zero only in a small volume fraction.
For this reason, the resulting violation of
energy remains negligible relative to the total energy budget.

In order to determine the parameter $C_{\mathrm{A}}$, we observe that
the Sharp-Wheeler SGS model is obtained as an asymptotic
relation in the limiting case of neglecting the non-local transport $\mathfrak{D}_{\mathrm{sgs}}$,
the turbulent energy transfer $\Sigma_{\mathrm{sgs}}$ and the pressure-dilatation
$\lambda_{\mathrm{sgs}}$ (see Sect.~2 of paper I for definitions). Dropping
the corresponding terms in equation~(\ref{eq:q_sgs}), one obtains
\begin{equation}
	\frac{\DD}{\DD t}q_{\mathrm{sgs}} \simeq
	\frac{1}{\sqrt{2}}C_{\mathrm{A}}g_{\mathrm{eff}} -
	\frac{q_{\mathrm{sgs}}^{2}}{\ell_{\epsilon}}.
\end{equation}
for a fluid parcel in the vicinity of the flame front. In the stationary regime,
this equation has the fixed point solution:
\begin{equation}
	q_{\mathrm{sgs}} \simeq 
	\sqrt{\frac{2C_{\mathrm{A}}\Delta_{\mathrm{eff}}g_{\mathrm{eff}}}{C_{\epsilon}}}
	=v_{\mathrm{RT}}(\Delta_{\mathrm{eff}}).
\end{equation}
Consistency with the Sharp-Wheeler relation~(\ref{eq:vel_sharp}) implies
$C_{\mathrm{A}}=C_{\epsilon}/8$. Since
$C_{\epsilon}\approx 0.5\ldots 1.0$ for developed turbulence
\citep[see][]{SchmHille05b}, the estimate $C_{\mathrm{A}}\approx 0.1$ is obtained.

\subsection{Rescaling of the subgrid scale turbulence energy}
\label{sc:rescaling}

If a non-static, co-moving grid is used, the implicit filter $\langle\ \rangle_{\mathrm{eff}}$
introduced in paper I, Sect.~3, becomes time-dependent. Therefore, time-derivates do not
commute with the operation of filtering and additional terms arise in the dynamical equations.
These terms are equivalent to the additional fluxes which are included in the implementation
of the Riemann solver for moving grids \citep{Mueller94,Roep05}. However, there is
a subtlety related to the shifting cutoff which separates the resolved and unresolved scales.
As the grid expands homologously with the bulk of the white dwarf, the grid resolution $\Delta$
gradually decreases in time and the growing cutoff length entails a gradual rise of the
SGS turbulence energy. This rise is inherent to the grid geometry and immediately affects the decomposition
of the energy budget. The two-thirds law for developed turbulence implies
$\langle q_{\mathrm{sgs}}\rangle\propto\Delta^{1/3}$ \citep[see][ Sect.~5]{Frisch}. 
Thus, it is easy to rescale the mean value of $q_{\mathrm{sgs}}$
if $\Delta$ changes by a small fraction $\delta\Delta/\Delta$:
\begin{equation}
	\langle q_{\mathrm{sgs}}^{(1+\delta)\Delta)}\rangle \propto
	\left(1+\frac{1}{3}\frac{\delta\Delta}{\Delta}\right)\langle q_{\mathrm{sgs}}^{\Delta}\rangle
\end{equation}

Because this scaling law is a statistical rule one cannot expect that it holds locally.
However, the dynamical procedure for the calculation of the eddy-viscosity
also can be utilised for a localised rescaling law. Let $\Delta_{t}$ be the grid
resolution at time $t$, and $\Delta_{t-\delta t}$ the slightly smaller
resolution prior to the last time step. Applying the test filter introduced in paper I, Sect.~4.1, at time $t$,
the turbulent velocity $q_{\mathrm{T}}$ associated with velocity fluctuations
on length scales greater than $\beta\Delta_{t}$ and smaller than $\gamma_{\mathrm{T}}\beta\Delta_{t}$
is obtained. Here $\beta=\Delta_{\mathrm{eff}}/\Delta$ is the constant ratio of the
effective cutoff length to the size of the grid cells. 
Since the Riemann solver does not account for the fractional
growth of the turbulence energy due to the shift of the cutoff, the
result of advancing the dynamical equations from $t-\delta t$ to $t$
is $q_{\mathrm{sgs}}^{\Delta_{t-\delta t}}$. Now an estimate of 
$q_{\mathrm{sgs}}^{\Delta_{t}}$ can be made locally via interpolation
of the turbulence energy in length scale space. 
Using the contracted Germano identity (paper I, Sect.~4.1), 
the total turbulence
energy associated with the test filter length at time $t$ is given by
\begin{equation}
	k^{\gamma_{\mathrm{T}}\Delta_{t}}_{\mathrm{turb}} =
    \frac{\langle\rho k_{\mathrm{sgs}}^{\Delta_{t}}\rangle_{\mathrm{T}}}{\rho^{(\mathrm{T})}}+
    k^{(\mathrm{T})}.
\end{equation}
The unknown is the rescaled SGS turbulence energy $k_{\mathrm{sgs}}^{\Delta_{t}}$.
Neglecting compressibility effects and variations on sub-test filter lengths,
we set $k^{\gamma_{\mathrm{T}}\Delta_{t}}_{\mathrm{turb}}\simeq
k_{\mathrm{sgs}}^{\Delta_{t}}+k^{(\mathrm{T})}$. Then linear interpolation
between $\Delta_{t-\delta t}$ and $\gamma_{\mathrm{T}}\Delta_{t}$ yields:
\begin{equation}
	\label{eq:vel_sgs_rescl_lin}
	k_{\mathrm{sgs}}^{\Delta_{t}} \simeq 
	k_{\mathrm{sgs}}^{\Delta_{t-\delta t}} + \frac{f}{1-f} k_{\mathrm{T}},
\end{equation}
where the interpolating factor $f$ is given by
\begin{equation}
	f = \frac{\Delta_{t}-\Delta_{t-\delta t}}{\gamma_{\mathrm{T}}\Delta_{t}-\Delta_{t-\delta t}}.
\end{equation}
Due to the smallness of a CFL time step, the fractional changes of the cutoff length
will be small. Hence, $f\ll 1$.

The problem with the rescaling law~(\ref{eq:vel_sgs_rescl_lin}) is that it fails to
account for the correct asymptotic behaviour in the limit of fully developed
turbulence. On account of the Germano identity (paper I, Sect.~4.1), 
one would expect the statistical relation
\begin{equation}
	(\gamma_{\mathrm{T}}^{2/3}-1)\langle k_{\mathrm{sgs}}\rangle=
	\langle k^{(\mathrm{T})}\rangle
\end{equation}
for regions of nearly homogeneous turbulence
obeying Kolmogorov scaling. This relation is asymptotically
reproduced by the rescaling modified law
\begin{equation}
	k_{\mathrm{sgs}}^{\Delta_{t}} \simeq 
	\frac{\gamma_{\mathrm{T}}^{2/3}(1-f)}{\gamma_{\mathrm{T}}^{2/3}-f} k_{\mathrm{sgs}}^{\Delta_{t-\delta t}} + 
	\frac{f}{\gamma_{\mathrm{T}}^{2/3}-f} k_{\mathrm{T}},
\end{equation}
which results from the interpolation of the turbulence energy divided
by the associated length to the power $2/3$. This is the rescaling law
which was implemented for the numerical simulations of thermonuclear
supernova explosions discussed in the next section.

\section{Numerical simulations}
\label{sc:num_simul}

In the following, we will present results from several numerical
simulations using the methodology outlined in \citet{RoepHille05}.  In
essence, the piece-wise parabolic method is used to solve the
hydrodynamical equations \citep{ColWood84,FryMueller89} and the
evolution of the flame-fronts is computed by means of the level set
method in the passive implementation \citep{OshSeth88,ReinHille99a}.
The implemented equation of state for electron-degenerate matter is
described in \citet{Rein01}, Sect.~3.2.  Thermonuclear burning is
modelled by simple representative reactions \citep{ReinHille02a}:
${^{12}}$C and ${^{16}}$C is fused to ${^{56}}$Ni and
$\alpha$-particles at densities higher than $5.25\cdot
10^{7}\,\mathrm{g\,cm^{-3}}$, whereas ${^{24}}$Mg is produced at lower
densities in the late stage of the explosion. Finally, all reactions
cease below $10^{7}\,\mathrm{g\,cm^{-3}}$.  This threshold presumably
marks the transition from the flamelet to the broken reaction zone
regime of turbulent burning \citep[see][]{NieKer97}.  The correct
treatment of the burning process in this regime is still a matter of
debate and, for this reason, not included in the present simulations.

As initial model, we choose a white dwarf of mass $M=1.4M_{\sun}$
composed of equal mass fractions of carbon and oxygen with a central density
$\rho_{\mathrm{c}}=2.0\cdot 10^{9}\,\mathrm{g\,cm^{-3}}$ and temperature
$T_{\mathrm{c}}=7.55\cdot 10^{8}\,\mathrm{K}$. As suggested by \citet{WunWoos04},
the radial temperature profile is given by a parabola with a cutoff
at the thermal radius $\Lambda=7.35\cdot 10^{7}\,\mathrm{cm}$:
\begin{equation}
	T(r) = T_{\mathrm{c}}\left[1-\left(\frac{r}{\Lambda}\right)^{2}\right]\theta(r-\Lambda) + 
	       T_{0}\theta(\Lambda-r),
\end{equation}
where $\theta$ denotes the Heaviside step function. 
The thermal radius specifies the size of the convective core prior to 
the run-away. 
At larger radii, the matter is isothermal with $T_{0}=5\cdot 10^{5}\,\mathrm{K}$.
In the centre, we set an axisymmetric initial burning region
with sinusoidal perturbations \citep[see][]{RoepHille05}.
In order to achieve higher resolution, only single octants
subject to reflecting boundary conditions were evolved 
in the simulations discussed here.

Moreover, we applied the co-expanding grid technique of \citet{Roep05}. Thereby,
it is possible to maintain an equidistant grid geometry over the whole domain of turbulent
burning at any stage of the explosion, even when the ejecta have expanded by
a large factor compared to the initial size of the white dwarf.
Recently, \citet{RoepHille05b} have combined this technique with the
grid geometry used by \citet{ReinHille02a} in order to capture the burning process in the 
interior with optimal resolution, while using a coarser grid with exponentially 
increasing cells outside.
The hybrid grid, even at moderate resolution, enables us to resolve details of the 
turbulent flame dynamics which used to be inaccessible for non-adaptive schemes.
All numerical simulations presented in this article feature a hybrid grid.

\begin{figure*}[thb]
  \begin{center}
    \includegraphics[width=17cm]{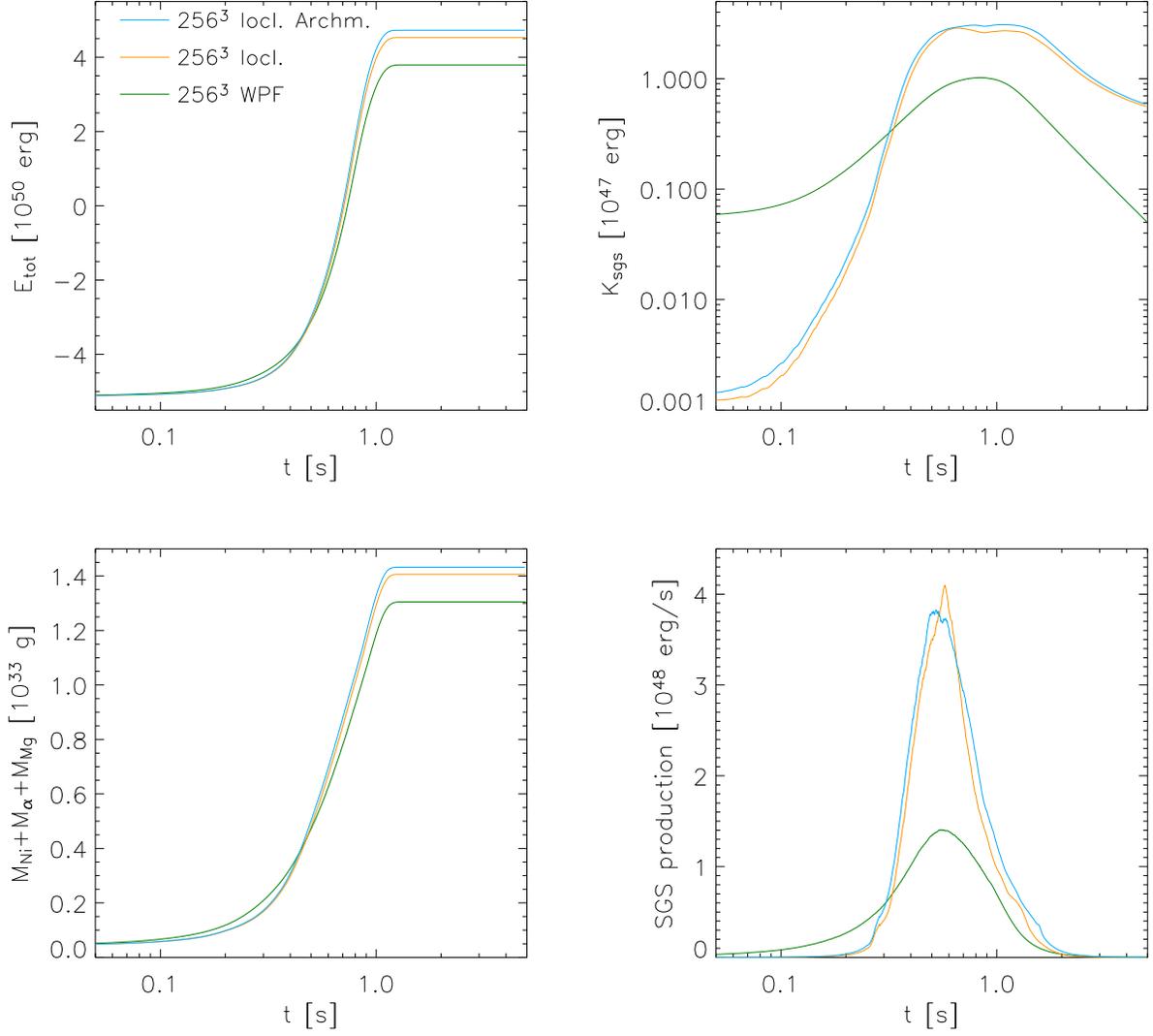}
    \caption{Time evolution of integrated quantities for three simulations with
    identical initial conditions and resolution $256^{3}$. In one case, Clement's
    SGS model with wall proximity functions (WPF) was used. For the other two
    simulations we applied the localised SGS model with
    Archimedian production, respectively, switched off and on.}
    \label{fg:evol_256}
  \end{center}
\end{figure*}

\begin{figure*}[thb]
  \begin{center}
    \includegraphics[width=17cm]{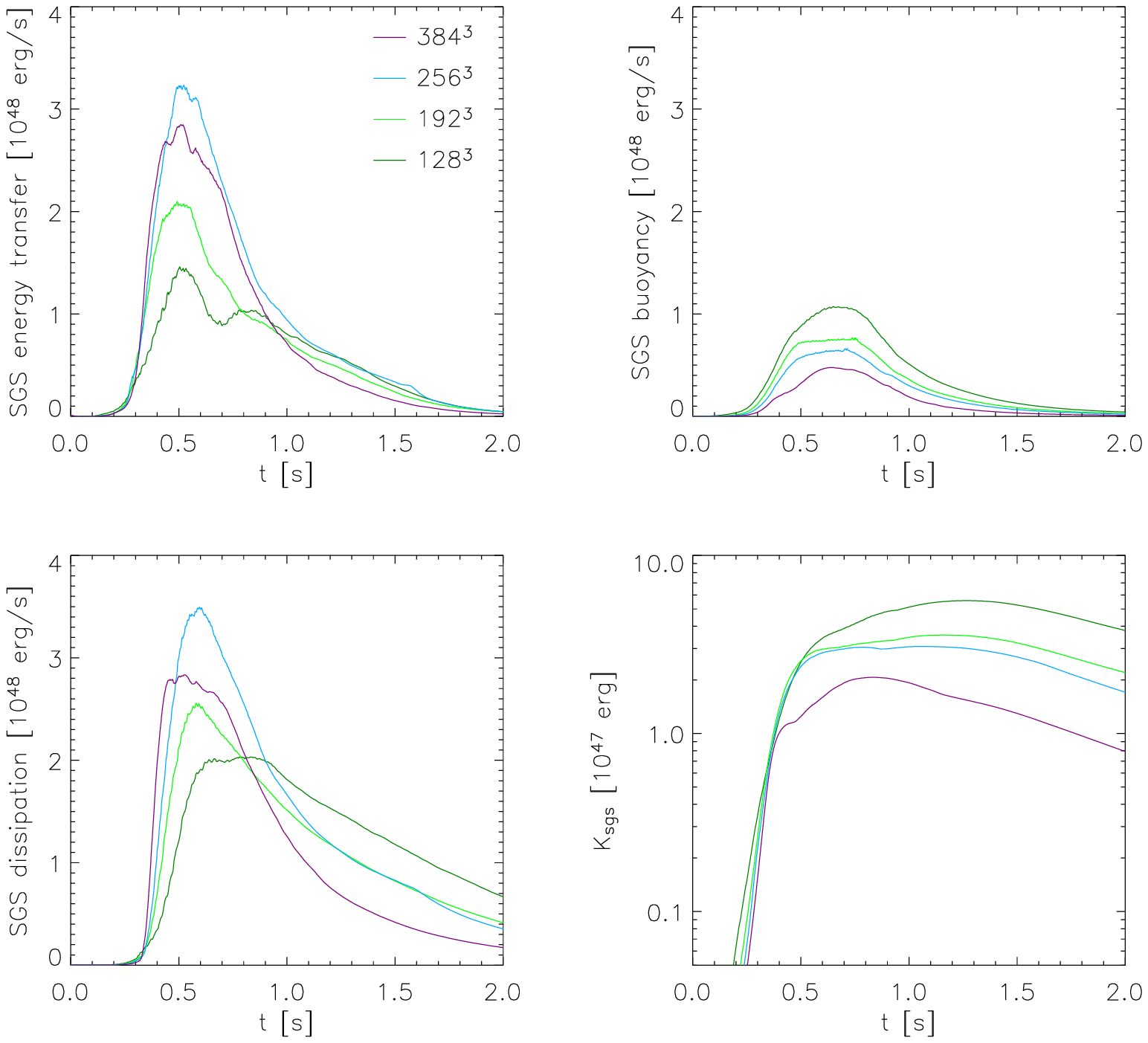}
    \caption{Time evolution of integrated SGS quantities for a series of simulations
    	with varying resolution. }
    \label{fg:evol_sgs_res}
  \end{center}
\end{figure*}

The non-uniform grid geometry poses certain difficulties when
applying the localised SGS model. In Sect.~\ref{sc:rescaling}, we showed that
it is relatively easy to account for the variation in the time domain
due the co-expansion of the grid. 
In the case of non-uniform grids, however, the filter operation does not
commute with spatial derivatives in the dynamical equations. Apart from that,
the weighing of nodes for the discrete filtering procedure becomes dependent on the
location. This would lead to substantial complications in the 
numerical implementation. Fortunately, we found a simple solution:
Since the turbulent dynamics mostly takes place in the burning region which
is contained within the uniform part of the grid, we computed the eddy-viscosity
parameter $C_{\nu}$ only in this region and set the rate of turbulent energy transfer
equal to zero in the exterior. Later in this section, we will demonstrate 
that neglecting the energy transfer outside the burning region
can be justified \emph{a posteriori}.
  
The original SGS turbulence energy model implemented by
\citet{NieHille95a} is based on statistical closure parameters. 
\citet{Clement93} suggested to set $C_{\nu}=0.1W$, 
where $W$ is an empirical \emph{wall proximity function} (WPF),
\begin{equation}
  \label{eq:wall_prox}
  W =
  \min\left[100,
    \max\left(0.1,\cdot 10^{-4}\frac{e_{\mathrm{int}}}{k_{\mathrm{sgs}}}\right)
  \right].
\end{equation}
Since $e_{\mathrm{int}}\sim c_{\mathrm{s}}^{2}$, the ratio
$e_{\mathrm{int}}/q_{\mathrm{sgs}}^{2}$ is 
basically an inverse Mach number squared. If there
is little turbulence energy, $C_{\nu}$ is considerably enhanced. On
the other hand, $C_{\nu}$
becomes smaller than $0.1$ if the SGS turbulence velocity exceeds a few percent
of the speed of sound. This behaviour of the eddy-viscosity parameter is 
qualitatively different from the prediction of the dynamical procedure
which implies less energy transfer if turbulence is still developing
but enhanced transfer in the fully developed case \citep{SchmHille05b}.

This is clearly reflected in the time evolution of the turbulence
energy in two simulations which differ only in the SGS model. The
graphs of the mass-integrated SGS turbulence energy are plotted in the
top panel on the left of Fig.~\ref{fg:evol_256}. There are two
variants of the localised SGS model, one including Archimedian
production as described in Sect.~\ref{sc:archm_prod}, whereas it is
assumed that energy transfer through the cascade is the only source of
turbulence production in the alternative model. In contrast to
Clement's model, there is initially very little SGS turbulence energy
followed by a much steeper rise in the simulations with the localised
SGS model.  The rapid growth of turbulence energy can be attributed to
the substantially stronger turbulence production within the time
interval from $0.3$ to $1.0\,\mathrm{s}$ (see right bottom panel in
Fig.~\ref{fg:evol_256}). In the second half of the combustion phase,
the total SGS turbulence energy is almost one order of a magnitude
larger which enhances the flame propagation speed accordingly.  At
later times, the discrepancy becomes even more pronounced because the
rescaling of $k_{\mathrm{sgs}}$ implemented in the localised model
feeds kinetic energy into the subgrid scales against the action of SGS
dissipation. The net result is a significant enhancement of the
explosion energy and a larger yield of burning products (see bottom
panel on the left of Fig.~\ref{fg:evol_256}). Also note that the
additional production of SGS turbulence by the Archimedian force term
in equation~(\ref{eq:q_sgs}) increases the explosion energy even
further.  Compared to the reference simulation with Clement's model,
the final kinetic energy of $0.472\cdot 10^{51}\,\mathrm{erg}$ is
about $25\,\%$ greater. However, this does not imply that Archimedian
production dominates over the turbulence cascade. In fact, the
evolution of the SGS turbulence energy differs only little, as one can
see from the plot in Fig.~\ref{fg:evol_256}.  In conclusion, SGS
buoyancy effects appear to influence the explosion but turbulent
energy transfer from resolved scales is nevertheless the primary
source of SGS turbulence production.

According to the scaling argument mentioned in the introduction,
buoyancy should become even less important relative to the turbulent
energy transfer with increasing resolution. This is indeed observed in
a series of simulations with the resolution varying between $128^{3}$
up to $384^{3}$ grid cells. The time evolution of the integrated rate
of energy transfer and specific Archimedian force, respectively, is
plotted in the top panels of Fig.~\ref{fg:evol_sgs_res}. In order to
interpret these graphs, it is important to note that the SGS energy
transfer is expected to become statistically scale invariant in the
case of Kolmogorov turbulence. This follows from the scaling law
$v'(l)\propto l^{1/3}$ for the turbulent velocity fluctuations.
Hence, $k_{\mathrm{sgs}}\propto \Delta^{2/3}$. Since the
characteristic time scale of turbulent velocity fluctuations scales
with $l^{2/3}$, it follows that the average time derivative of
$k_{\mathrm{sgs}}$ is scale invariant. The rate of energy transfer per
unit mass, on the other hand, is proportional to $\Delta
k_{\mathrm{sgs}}^{1/2}|S^{\ast}|^{2}$. This expression is also
scale-invariant because the rate-of-strain scalar $|S^{\ast}|$
measures the inverse time scale of the smallest resolved velocity
fluctuations, i.e.  $|S^{\ast}|^{2}\propto\Delta^{-4/3}$, while
$\Delta k_{\mathrm{sgs}}^{1/2}\propto\Delta^{4/3}$.  The computed rate
of energy transfer plotted in the top panel on the left of
Fig.~\ref{fg:evol_sgs_res} exhibit peak values which are within the
same order of magnitude, although the energy transfer seems to be
underestimated for the lowest resolutions.  Initially, the
turbulence energy rises exponentially at a rate which changes only
little with resolution (right bottom panel in
Fig.~\ref{fg:evol_sgs_res}). This is reflected in the nearly
coinciding graphs of the rate of SGS energy transfer up to $t\approx
0.3\,\mathrm{s}$ shown in the left panel on the top of
Fig.~\ref{fg:evol_sgs_res}.

\begin{figure}[thb]
  \begin{center}
    \resizebox{\hsize}{!}{\includegraphics{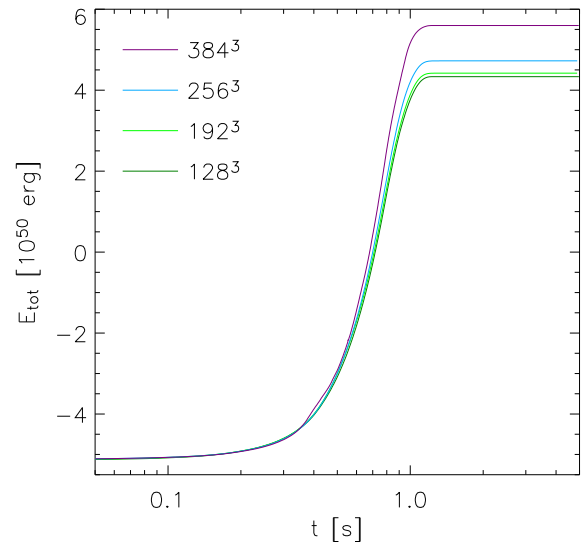}}
    \caption{Time evolution of the total energy for the same simulations
             as in Fig.~\ref{fg:evol_sgs_res}}
    \label{fg:evol_energy}
  \end{center}
\end{figure}

The SGS turbulence energy in the regime of fully developed turbulence
decreases for higher resolution. This trend can be discerned
particularly in the post-burning phase, in which no further energy is
injected and the turbulent flow begins to decay. From the initial
production phase to the post-burning phase, however, a rather
complicated behaviour becomes manifest as the result of the interplay
between turbulence production by the strain of the resolved flow, the
SGS Archimedian force and the grid expansion.  The explosion
energetics plotted in Fig.~\ref{fg:evol_energy} shows the following
behaviour depending on the numerical resolution (also see
Table~\ref{tb:evol_burn_res}): Whereas the final value of the total
energy is about the same for the lower resolutions, there is a
significantly enhanced yield of energy in the case
$N=384^{3}$. However, this does not imply that the model fails to
converge with increasing resolution. Turbulent flow regions are
confined in a fraction of the numerical grid, whereas the greater part
of the grid is overhead required for modelling the non-turbulent outer
parts of the expanding star and some portion of the surrounding
quasi-vacuum. In the case $N=256^{3}$, for example, significant SGS
turbulence production occurs in the inner $100^{3}$ cells at
$t=0.5\,\mathrm{s}$. This is the time of maximal integrated turbulence
production. However, in paper I we demonstrated that $100$ cells in
each spatial dimension is definitely not sufficient to resolve
developed turbulent flow sufficiently far down toward the inertial
subrange using PPM. Although this is merely a crude estimate, it
appears plausible that even the supernova simulation with $N=384^{3}$
resolves the turbulent dynamics only marginally. As a further
indication, the plateau-like flattening of the rate of production and
dissipation, respectively, can be seen in the left panels of
Fig.~\ref{fg:evol_sgs_res} for the highest resolution only. We
interpret the flattening as a consequence of local statistical
equilibrium between production and dissipation.

Unfortunately, we were not able to perform a run of still higher
resolution due to the limitations of our computational resources. In
any case, we expect that $N=512^{3}$ grid cells in one octant would be
sufficient for an accurate modelling of the turbulent dynamics,
whereas simulations with less resolution can be utilised to discern
trends in parameter studies. A similar conclusion was drawn by
\citet{Roep05} from a series of two-dimensional simulations, in which
a pronounced jump of the total energy was found between the
$N=256^{2}$ and the $512^{2}$ run, respectively, while more or less
the same energy was obtained for $N\ge 512^{2}$. Moreover, snapshots
of the zero level set for varying resolution suggested that secondary
Kelvin-Helmholtz instabilities are barely or not at all resolved with
$N\le 256^{2}$.

\begin{table}[htb]
  \caption{Total release of nuclear and kinetic energy and
           total masses of iron group (``Ni'') and intermediate mass
           (``Mg'') elements corresponding to
           Fig.~\ref{fg:evol_energy}.}
  \label{tb:evol_burn_res}
  \begin{center}
    \begin{tabular}{c c c c c}
      \hline
      $N$ & $E_{\mathrm{nuc}}\,[10^{51}\,\mathrm{erg}]$ &  
      $E_{\mathrm{kin}}\,[10^{51}\,\mathrm{erg}]$ &
      $M_{\mathrm{Ni}}/M_{\sun}$ & $M_{\mathrm{Mg}}/M_{\sun}$ \\
      \hline\hline
      $128^{3}$ & 0.963 & 0.433 & 0.523 & 0.175 \\
      $192^{3}$ & 0.970 & 0.442 & 0.529 & 0.172 \\
      $256^{3}$ & 1.000 & 0.472 & 0.548 & 0.172 \\
      $384^{3}$ & 1.087 & 0.560 & 0.586 & 0.206 \\
      \hline
    \end{tabular}
  \end{center}
\end{table}

\begin{figure*}[thb]
  \begin{center}
    \includegraphics[width=17cm]{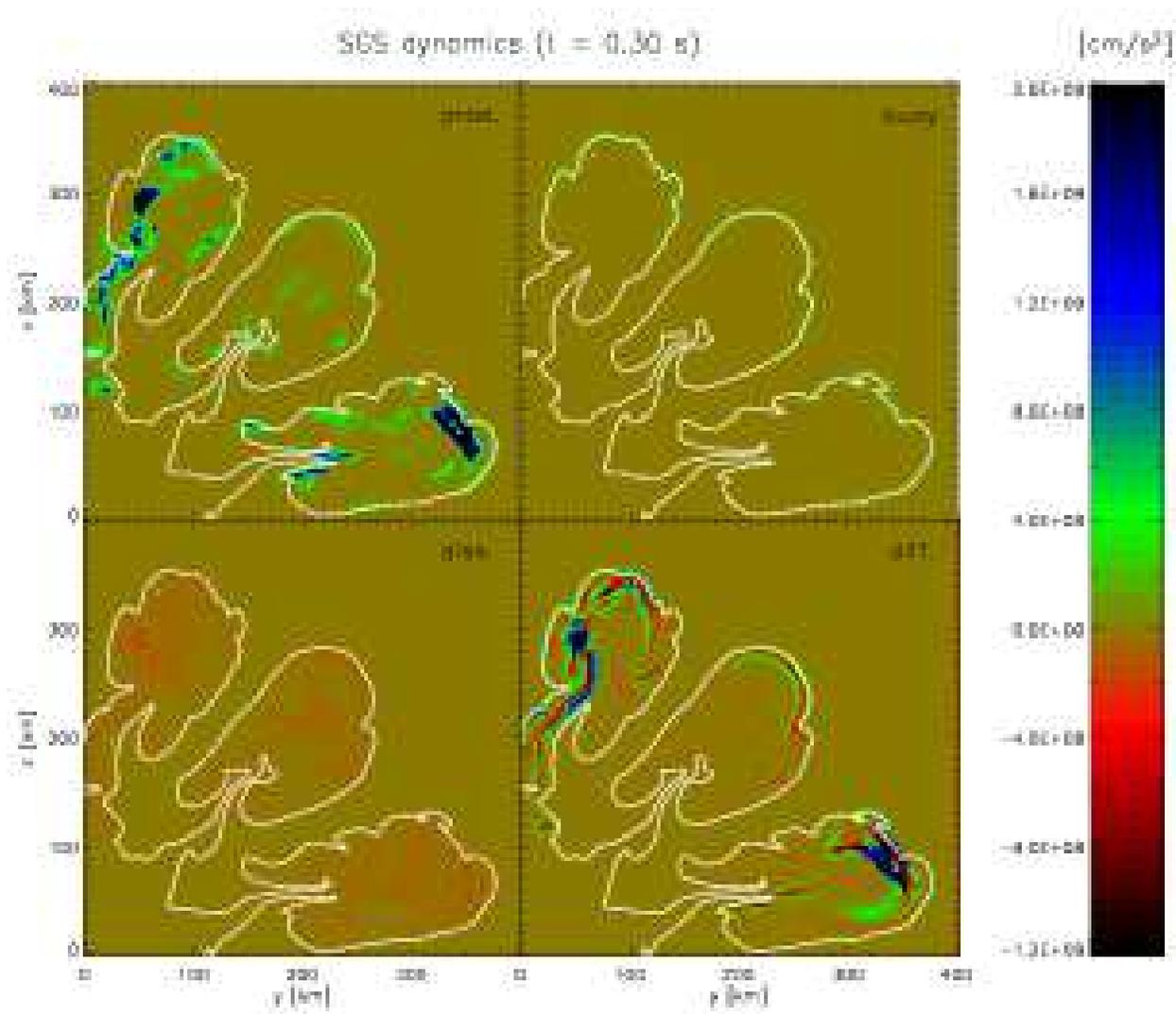}
    \caption{Contour sections showing the contributions to the evolution of
    	the SGS turbulence velocity $q_{\mathrm{sgs}}$ given by equation~(\ref{eq:q_sgs})
    	at $t=0.3\,\mathrm{s}$. Only the inner region of the grid with $N=384^{3}$ cells is
    	shown. The white contours represent the sections through the flame surface.}
    \label{fg:sgs_030}
  \end{center}
\end{figure*}

\begin{figure*}[thb]
  \begin{center}
    \includegraphics[width=17cm]{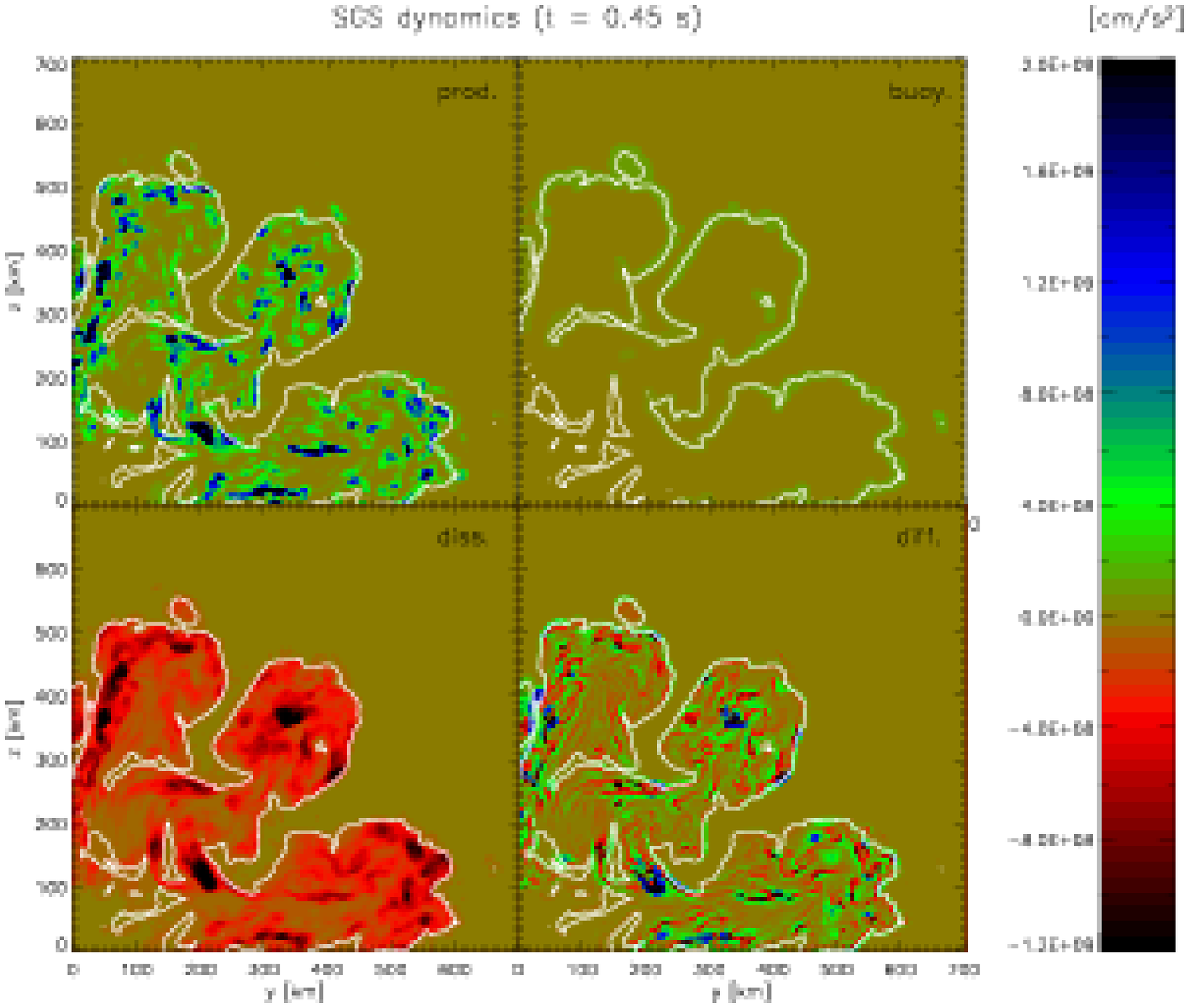}
    \caption{The same plot as in Fig.~\ref{fg:sgs_030} at $t=0.45\,\mathrm{s}$.}
    \label{fg:sgs_045}
  \end{center}
\end{figure*}

\begin{figure*}[thb]
  \begin{center}
    \includegraphics[width=17cm]{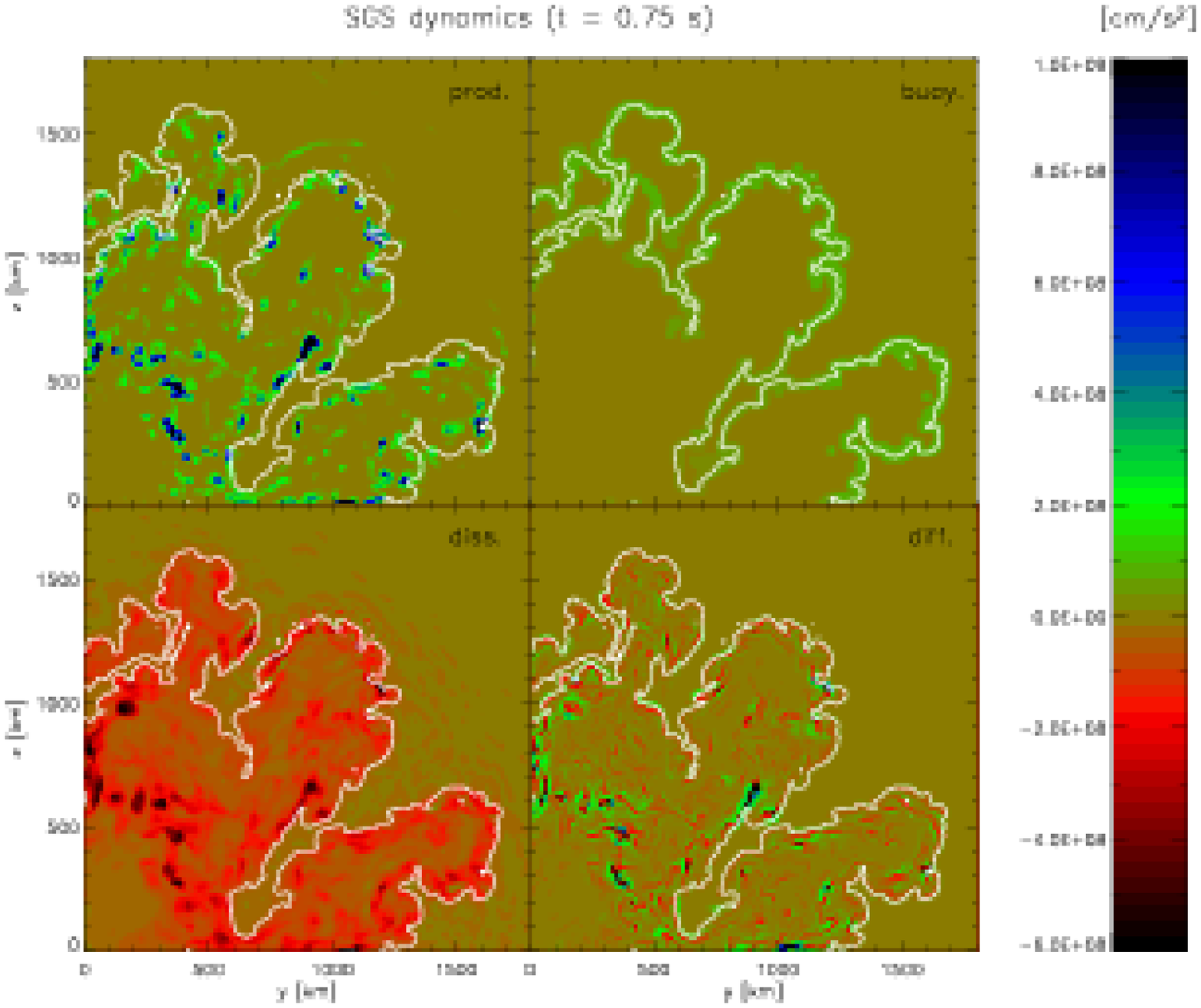}
    \caption{The same plot as in Fig.~\ref{fg:sgs_030} at $t=0.75\,\mathrm{s}$.}
    \label{fg:sgs_075}
  \end{center}
\end{figure*}

\begin{figure*}[thb]
  \begin{center}
    \includegraphics[width=17cm]{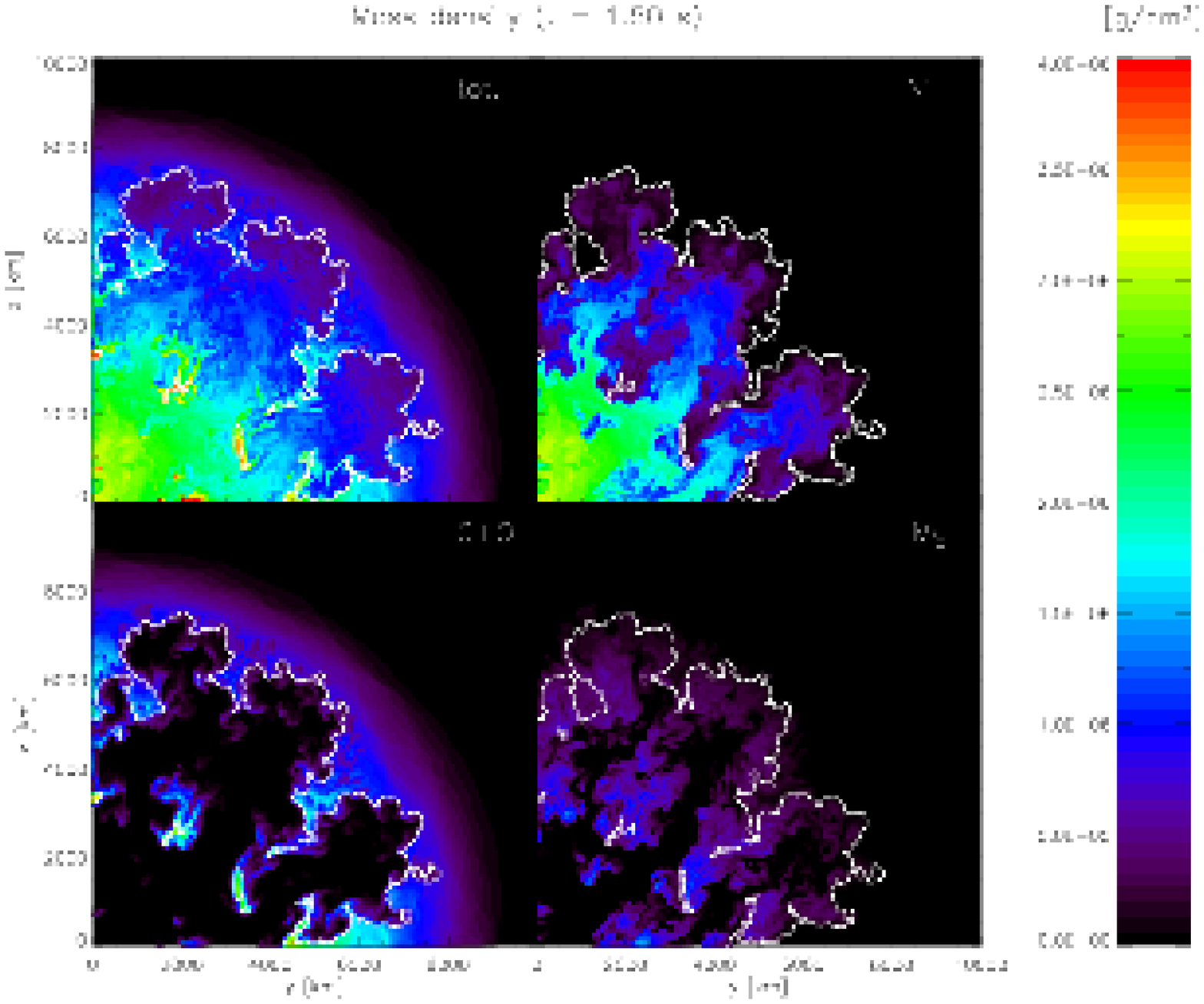}
    \caption{Total and partial mass densities for the same simulation as in
    	 at Fig.~\ref{fg:sgs_030}--\ref{fg:sgs_075} at $t=1.5\,\mathrm{s}$.}
    \label{fg:densty_150}
  \end{center}
\end{figure*}

Details of the SGS dynamics are illustrated by contour plots of
two-dimensional spatial sections from the simulation with $N=384^{3}$
grid cells (Figs.~\ref{fg:sgs_030}, \ref{fg:sgs_045}, and~\ref{fg:sgs_075}). 
In each Fig., the following dynamical terms of equation~(\ref{eq:q_sgs})
are plotted:
\begin{enumerate}
\item Rate of production caused by strain, $\ell_{\nu}|S^{\ast}|^{2}$ (left top panel).
\item Specific Archimedian force $0.1g_{\mathrm{eff}}$  (right top panel).
\item Rate of dissipation $- \frac{7}{30}q_{\mathrm{sgs}}d -
	q_{\mathrm{sgs}}^{2}/\ell_{\epsilon}$ (left bottom panel).
\item Rate of diffusion $\frac{1}{\rho}
	\vec{\nabla}\cdot\left(\rho\ell_{\kappa}q_{\mathrm{sgs}}
	\vec{\nabla}q_{\mathrm{sgs}}\right) -
	\ell_{\kappa}|\vec{\nabla}q_{\mathrm{sgs}}|^{2}$ (right bottom panel).
\end{enumerate}
The flame surface as given by the zero level set is indicated
by the contours in white.  Note that these quantities have the
dimension of acceleration. Fig.~\ref{fg:sgs_030} shows the typical
Rayleigh-Taylor mushroom shapes which have formed out of the initial
sinusoidal perturbations at time $t=0.3\,\mathrm{s}$. Significant
energy transfer is concentrated in small regions and there is little
dissipation yet. Comparing to Fig.~\ref{fg:evol_sgs_res}, one can see
that turbulence production is just about to rise. At
$t=0.45\,\mathrm{s}$, the rate of energy transfer has reached its
maximum and is spread all over the interior of the flames (see
Fig.~\ref{fg:sgs_045}).  The acceleration of SGS fluid parcels subject
to the largest strain exceeds $10^{6}$ times the gravitational
acceleration on Earth relative to the resolved flow.  The SGS buoyancy
is typically by an order of a magnitude smaller. Both dissipation and
transport due to SGS turbulent diffusion are comparable to the rate of
energy transfer at this time.  In the unburned material outside, on
the other hand, there is virtually no SGS turbulence.  Thereby, it is
confirmed that switching off the energy transfer terms in the
non-uniform gird regions at sufficient distance from the flame fronts
is a reasonable simplification.  Obviously, the flow is highly
anisotropic in the vicinity of the flames which highlights the
necessity of a localised SGS model. In Fig.~\ref{fg:sgs_075}, one can
see that turbulent energy transfer is declining and becoming small
relative to the Archimedian force near the flame front. However, this
does not imply that the amount of SGS turbulence and, thus, the
turbulent flame speed is dominated by the SGS buoyancy because the
bulk of SGS turbulence energy has been produced by transfer of kinetic
energy through the turbulence cascade and diffusion acts to
redistribute this energy into regions with little production.

\begin{figure}[htb]
  \begin{center}
    \resizebox{\hsize}{!}{\includegraphics[width=17cm]{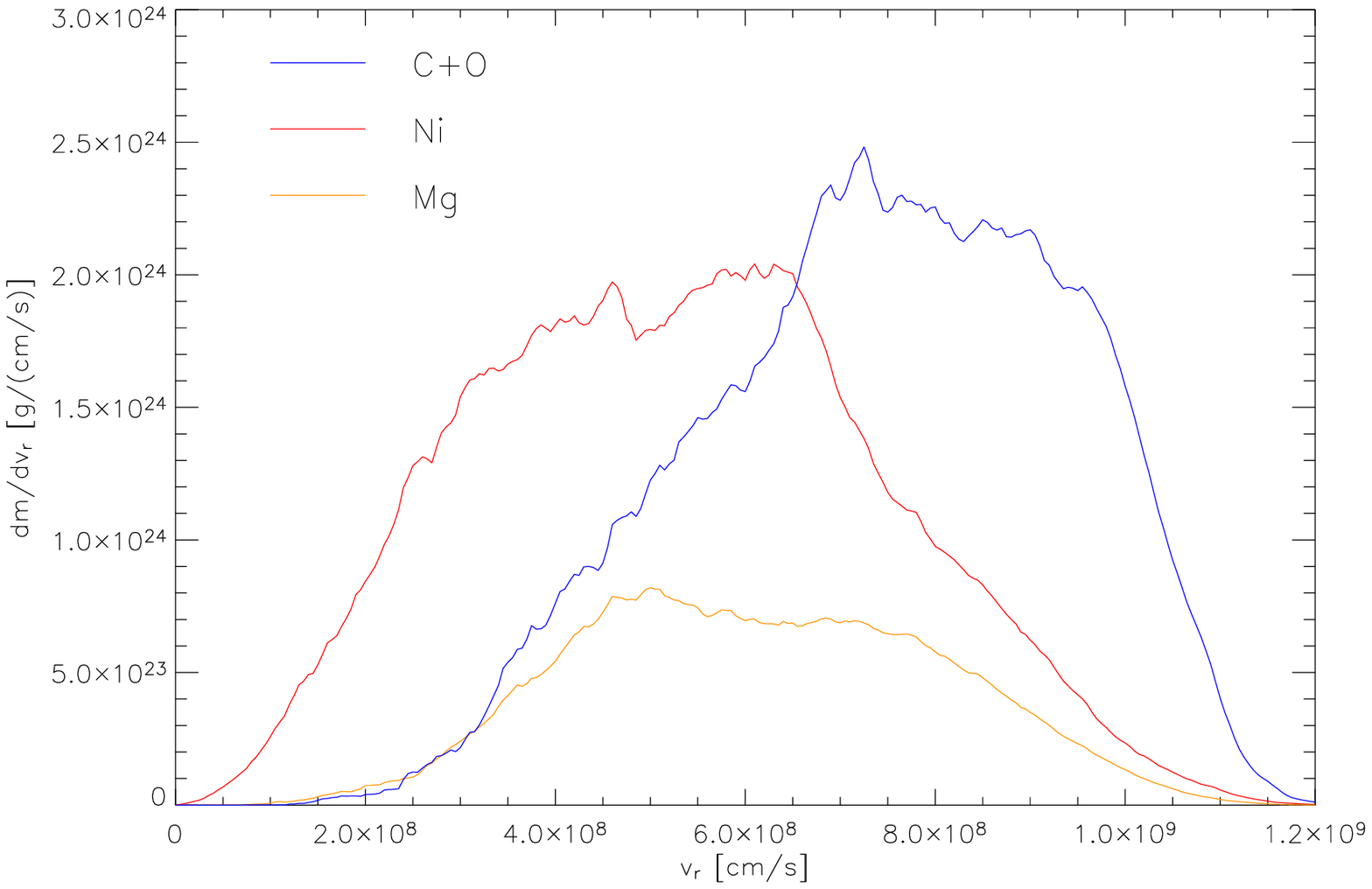}}
    \caption{Density functions of mass in radial velocity space for the major nuclear species
      in the same simulation as in Fig.~\ref{fg:sgs_030}--\ref{fg:densty_150} at $t=5.0\,\mathrm{s}$.}
    \label{fg:mass_distrb_500}
  \end{center}
\end{figure}

\begin{figure}[thb]
  \begin{center}
    \resizebox{\hsize}{!}{\includegraphics[width=17cm]{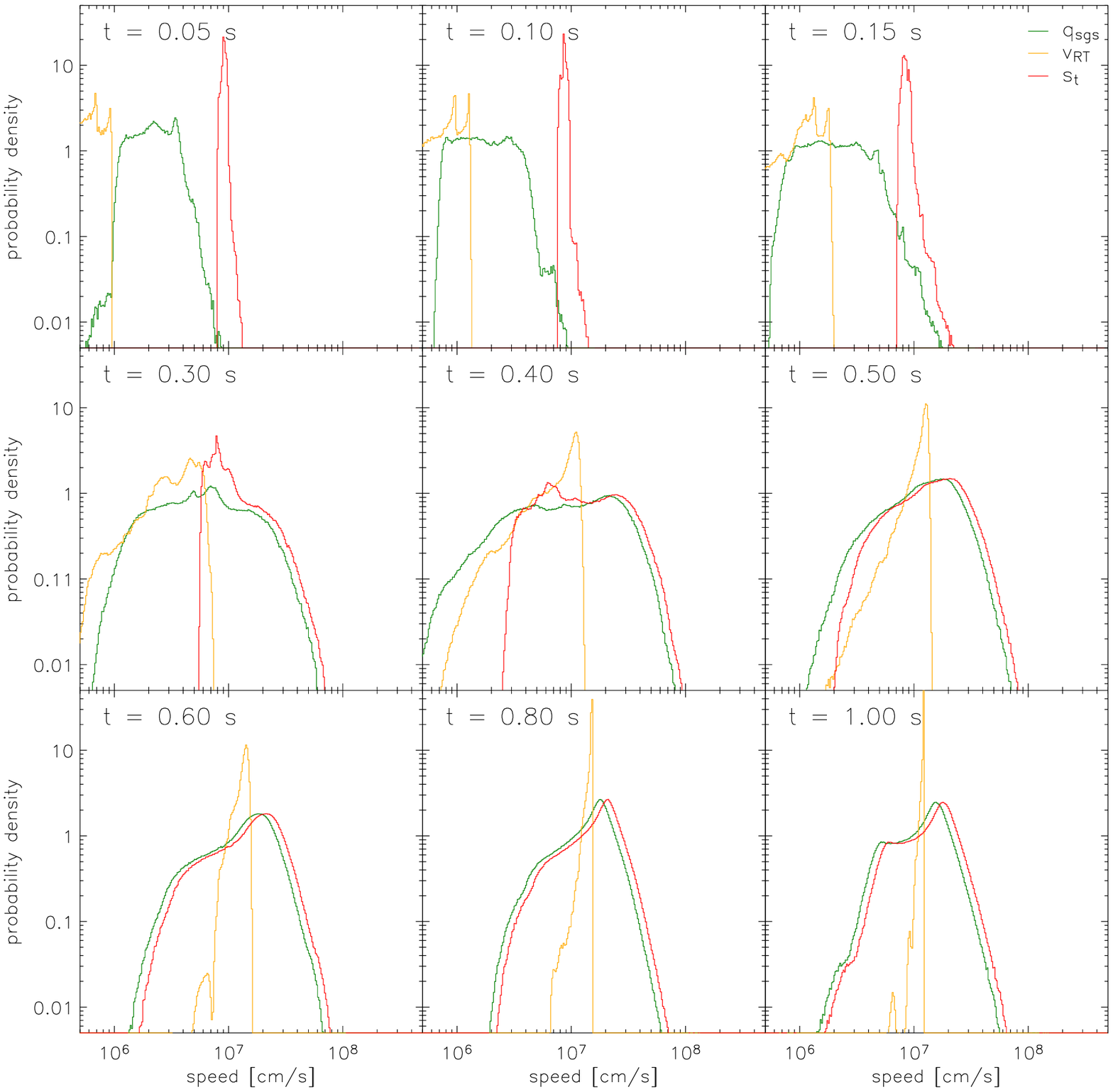}}
    \caption{Probability density functions of $v_{\mathrm{RT}}(\Delta_{\mathrm{eff}})$, $q_{\mathrm{sgs}}$
    	and $s_{\mathrm{t}}$ over the flame surface at several instants of time.}
    \label{fg:pdf_speed}
  \end{center}
\end{figure}

Fig.~\ref{fg:c3_3d} shows the evolution of the SGS turbulence velocity $q_{\mathrm{sgs}}$ 
at the flame fronts in three-dimensional visualisations. The grid
lines roughly indicate
the uniform part of the numerical grid. The corresponding absolute scale is indicated by the
size $X_{\mathrm{uni}}$ and the corresponding number of cells $N_{\mathrm{uni}}$.
In the first three snapshots
one can see the growth of the initial perturbations. The axial
symmetry is gradually broken
by the formation of secondary instabilities. At $t=0.45\,\mathrm{s}$ the smaller plumes
originating from these instabilities are highly turbulent. From $t=0.6\,\mathrm{s}$
onwards, the system increasingly looses its memory of the initial condition and the
turbulence intensity at the flame surface is abating and levelling. The last snapshot
at $t=1.5\,\mathrm{s}$ shows a complex structure with features over a wide range of
scales. There appear to be five or six major modes which eventually prevail.
The resulting layering of nuclear species is illustrated by the contour plots
of the corresponding mass densities in
Fig.~\ref{fg:densty_150}. Nickel is concentrated in the central region, 
whereas both magnesium and unburned carbon and oxygen are
found further outside. The outermost layers and the narrow down-drafts between
the convective fingers of nuclear ash are composed almost exclusively of carbon and oxygen.
The stratification of the nuclear species in the explosion ejecta is reflected
in the corresponding mass density functions $\dd M/\dd v_{r}$, where $v_{\mathrm{r}}$
is the radial velocity component.  In particular, Fig.~\ref{fg:mass_distrb_500} shows that
little carbon and oxygen is found for velocities less than $3000\,\mathrm{km\,s^{-1}}$
in the late phase of almost homologous expansion.

To understand the flame dynamics, it is instructive to consider the
probability density function (PDF) of the logarithm of the flame
propagation speed $s_{\mathrm{t}}$ over the surface of the flame.  The
PDFs for several instants of time are plotted in
Fig.~\ref{fg:pdf_speed}. Note that integrating each PDF over the
decade logarithm of the speed yields unity. Also shown are the PDFs of
$q_{\mathrm{sgs}}$ and $v_{\mathrm{RT}}(\Delta_{\mathrm{eff}})$. The
relation between the turbulent flame speed $s_{\mathrm{t}}$ and the
SGS turbulence velocity $q_{\mathrm{sgs}}$ is formulated in
equation~(\ref{eq:sgs_flame_speed_pocheau}).  During the first tenth
of a second, $s_{\mathrm{t}}$ is basically given by the laminar flame
speed. Then the flame propagation becomes increasingly affected by SGS
turbulence. From about $0.3\,\mathrm{s}$ onwards, $s_{\mathrm{t}}$ is
dominated by $q_{\mathrm{sgs}}$. At later times, one can see the
asymptotic relation $s_{\mathrm{t}}\simeq
2q_{\mathrm{sgs}}/\sqrt{3}$. The Rayleigh-Taylor velocity scale
$v_{\mathrm{RT}}(\Delta_{\mathrm{eff}})$ is initially much smaller
than the laminar burning velocity. As the flame propagates outwards,
both the gravity and the density contrast at the flame surface become
larger and $v_{\mathrm{RT}}(\Delta_{\mathrm{eff}})$ increases.
Eventually, the PDF of $v_{\mathrm{RT}}(\Delta_{\mathrm{eff}})$ tends
toward a rather narrow peak around $10^{7}\,\mathrm{cm\,s^{-1}}$. The
PDF of $q_{\mathrm{sgs}}$, on the other hand, extends over a
substantially wider range. For this reason, the localised SGS model
generates more variation in the propagation of the flame front in
comparison to the Sharp-Wheeler model.  This is expected because the
Sharp-Wheeler relation is ignorant of the interaction between subgrid
and resolved scales and the effects of non-local
transport. Nevertheless, $v_{\mathrm{RT}}(\Delta_{\mathrm{eff}})$ is
seemingly a velocity scale which is representative for the magnitude
of $q_{\mathrm{sgs}}$ at the flame surface during most of the burning
process.

\begin{figure*}[htb]
  \begin{center}
    \vspace{60mm}
    \mbox{\subfigure[$t=0.15\,\mathrm{s}$, $N_{\mathrm{uni}}=225^{3}$, $X_{\mathrm{uni}}=320\,\mathrm{km}$]{\texttt{\hspace{30mm}fig\_c3\_3d\_015.png\hspace{30mm}}}\qquad
          \subfigure[$t=0.3\,\mathrm{s}$, $N_{\mathrm{uni}}= 242^{3}$, $X_{\mathrm{uni}}=484\,\mathrm{km}$]{\texttt{\hspace{30mm}fig\_c3\_3d\_030.png\hspace{30mm}}}}\\
    \vspace{60mm}
    \mbox{\subfigure[$t=0.45\,\mathrm{s}$, $N_{\mathrm{uni}}= 258^{3}$, $X_{\mathrm{uni}}=721\,\mathrm{km}$]{\texttt{\hspace{30mm}fig\_c3\_3d\_045.png\hspace{30mm}}}\qquad
          \subfigure[$t=0.6\,\mathrm{s}$, $N_{\mathrm{uni}}=283^{3}$, $X_{\mathrm{uni}}=1260\,\mathrm{km}$]{\texttt{\hspace{30mm}fig\_c3\_3d\_060.png\hspace{30mm}}}}\\
    \vspace{60mm}
    \mbox{\subfigure[$t=0.75\,\mathrm{s}$, $N_{\mathrm{uni}}=308^{3}$, $X_{\mathrm{uni}}= 2190\,\mathrm{km}$]{\texttt{\hspace{30mm}fig\_c3\_3d\_075.png\hspace{30mm}}}\qquad
          \subfigure[$t=1.5\,\mathrm{s}$, $N_{\mathrm{uni}}=340^{3}$, $X_{\mathrm{uni}}=9120\,\mathrm{km}$]{\texttt{\hspace{30mm}fig\_c3\_3d\_150.png\hspace{30mm}}}}
    \caption{Evolution of the flames in the simulation with $N=384^{3}$ grid cells.
    	The colour shading indicates the value of $q_{\mathrm{sgs}}$ on a logarithmic scale.}
    \label{fg:c3_3d}
  \end{center}
\end{figure*}

\section{Conclusion}

We applied the SGS turbulence energy model to the large eddy
simulation of turbulent deflagration in thermonuclear supernova
explosions. The novel features of this model are a localised
closure for the rate of energy transfer, an additional
Archimedian force term which accounts for buoyancy effects on
unresolved scales and the rescaling of the SGS turbulence
energy due to the shift of the cutoff length in simulations with a
co-expanding grid. We found that the production of turbulence is
largely confined to the regions near the flame fronts and in the
interior ash regions. Consequently, there is pronounced anisotropy at
the flame surface which can be tackled by the localised SGS model
only.  The Archimedian force contributes noticeably to the turbulent
flame speed, particularly once the flame surface has grown
substantially. However, the dominating effect is the energy transfer
through the turbulence cascade. In the late stage of the explosion,
sustained turbulence energy comes from the rescaling, while the major
dynamical contribution is SGS dissipation. Furthermore, it
appears that numerical grids with more than $N=256^{3}$ cells in one
octant are necessary in order to sufficiently resolve the turbulent
dynamics in the burning regions and to obtain converged results.

An investigation of probability density functions over the flame
fronts (see Fig.~\ref{fg:pdf_speed}) reveals that the Rayleigh-Taylor
velocity scale $v_{\mathrm{RT}}(\Delta_{\mathrm{eff}})$ given by the
Sharp-Wheeler relation~(\ref{eq:vel_sharp}) is not negligible compared
to the SGS turbulence velocity $q_{\mathrm{sgs}}$, once the regime of
fully turbulent burning has been entered.  This reflects the slow
decrease of the ratio
$v_{\mathrm{RT}}(\Delta_{\mathrm{eff}})/q_{\mathrm{sgs}}$ with the
numerical cutoff scale according to the scaling argument discussed in
the introduction. The underlying scaling relations do not necessarily
apply to transient and inhomogenous flows as in the supernova
explosion scenario.  The PDF for $q_{\mathrm{sgs}}$ shows a
considerably wider spread than the sharply peaked PDF for
$v_{\mathrm{RT}}(\Delta_{\mathrm{eff}})$. We interpret this
observation as a consequence of the additional physics in the
localised SGS model, which also encompasses turbulent energy transfer
(i.e. interaction between resolved and subgrid scales) and turbulent
transport (i.e. non-local interactions among subgrid scales). The
relation between the Sharp-Wheeler and the turbulence energy models
may be analogous to the relation between the mixing length and
Reynolds stress models of convection.

The final kinetic energy in the simulation with the highest resolution
is about $6\cdot 10^{50}\,\mathrm{erg}$. The produced mass of iron
group elements, $0.58M_{\sun}$, falls within the range deduced from
observations of type Ia supernovae \citep{Leib00}. However, some
observed events are substantially more energetic. Regarding the
numerical simulations, the explosion energy is very sensitive to the
initial conditions and the localised SGS model appears to increase the
sensitivity even further. Using initial conditions that are different
to the highly artificial centrally ignited flame is in progress.  A
persistent difficulty is the large amount of left over carbon and
oxygen (see Fig.~\ref{fg:densty_150} and~\ref{fg:mass_distrb_500}). It
is not clear yet to what extent this problem can be solved with the
aid of the localised SGS model in simulations with more realistic
ignition scenarios. It appears more likely that a different mode of
burning is required in the late explosion phase. The currently
implemented numerical burning extinction at a density threshold of
$10^{7}\,\mathrm{g\,cm^{-3}}$ is mostly arbitrary and should be
replaced by a physical criterion motivated by the properties of
distributed burning at low densities. This might turn out to be an
alternative to the DDT scenario.

\begin{acknowledgements}
The numerical simulations were run on the Hitachi SR-8000 of the
Leibniz Computing Centre in Munich and the IBM p690 of the Computing
Centre of the Max-Planck-Society in Garching, Germany. The research of
W. Schmidt and J. C. Niemeyer was supported by the Alfried Krupp Prize
for Young University Teachers of the Alfried Krupp von Bohlen und
Halbach Foundation.
\end{acknowledgements}

\bibliographystyle{aa}

\bibliography{3618}

\end{document}